\documentclass{anabs}
\usepackage{graphicx}
\usepackage{psfig}
\usepackage{times}
\Volume{1-2}              
\Year{2003}              
\Month{02}               
\Pagespan{000}{000}      

\begin{document}
\lhead[\thepage]{A.N. Author: Title}
\rhead[Astron. Nachr./AN~{\bf 324} (2003) 1/2]{\thepage}
\headnote{Astron. Nachr./AN {\bf 324} (2003) 1/2, 000--000}

\title{Near-Infrared  
Identification of the Dimmest X-ray Sources on the Galactic Plane
with the ESO/NTT SOFI Camera}
\author{Ken Ebisawa\inst{1,2}, V. Beckmann\inst{1}, T. J.-L. Courvoisier\inst{1,3}, P. Dubath\inst{1}, 
 H. Kaneda\inst{4},  Y. Maeda\inst{4},  S. Yamauchi\inst{5} and E. Nishihara\inst{6}}
\institute{
INTEGRAL Science Data Centre, Chemin d'\'Ecogia 16, CH-1290 Versoix, Switzerland
\and 
Laboratory for High Energy Astrophysics, NASA/GSFC, Greenbelt, MD 20771, USA
\and 
Geneva Observatory, Chemin des Maillettes 51, CH-1290 Sauverny, Switzerland
\and 
Institute of Space and Astronautical Science, Yoshinodai 3-1-1, Sagamihara, Kanagawa 229-8510, Japan
\and 
Faculty of Humanities and Social Sciences, Iwate University,  Ueda 3-18-34, Morioka, Iwate 020-8550, Japan
\and 
Gunma Astronomical Observatory, 6860-86 Nakayama, Takayama-mura,  Agatsuma-gun, Gunma  377-0702, Japan}

\correspondence{ebisawa@obs.unige.ch}
\maketitle

\section{Introduction}
We have carried out a deep X-ray observation 
on a typical Galactic plane region with the Chandra ACIS-I instrument with 
unprecedented sensitivity and spatial
resolution (Ebisawa et al.\ 2001), and detected 274 unidentified  X-ray point sources
in the $\sim$  500 arcmin$^2$ region.
In order to identify these new X-ray sources, 
 we  have carried out a  near infrared follow-up observation 
using ESO/NTT (Tarenghi and Wilson 1989) SOFI  infrared camera 
on 2002/7/28 and 2002/7/29.

We have made a mosaic observation to cover the large Chandra field of view ($17' \times 17'$) with 
SOFI ($4.94'\times 4.94'$).
We have chosen 7 ``A'' fields and surrounding 9 ``B'' fields to cover
the central part of the Chandra field (Figure 1).
Exposure time  for $J$, $H$ and $K_s$ bands for an A-field is
10,  10 and  14 minutes, and those for a B-field
is    5,   5 and 7.47 minutes respectively.  The seeing was best in the first night ($\sim 0.6''$),
whereas in the second night  it was moderate ($\sim 1.5''$)
when we observed the remaining fields.

\section{Preliminary Results}

Quick look analysis suggests we have reached the limiting magnitude $\sim$ 20 mag 
for all the three bands in both A and B fields.
In Figure 2, we show a zoomed $K_s$-band image  of a part of  field A2.
The Chandra sources detected below $\sim 2$ keV are marked with red, and
those detected above $\sim 3 $ keV are marked with blue. All the five
soft X-ray sources in this field have near-infrared counterparts,
while the counterpart of the hard X-ray source (\#151) is not found.

Most of the soft X-ray sources are considered to be nearby active stars,
while hard X-ray sources are primarily background AGNs.
From the $J$, $H$ and $K_s$ band color analysis,
we plan to investigate the origin of these dim X-ray sources.

\begin{figure}
\centerline{
\psfig{figure=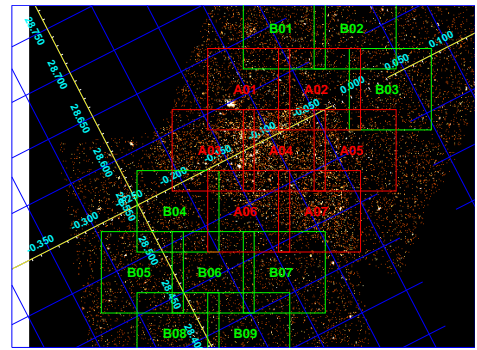,angle=0,width=6.2cm}
}
\caption{SOFI fields of view on the Chandra image.
Red ones are priority ``A'', while green ones are
priority ``B''.
}
\end{figure}

\begin{figure}
\centerline{
\psfig{figure=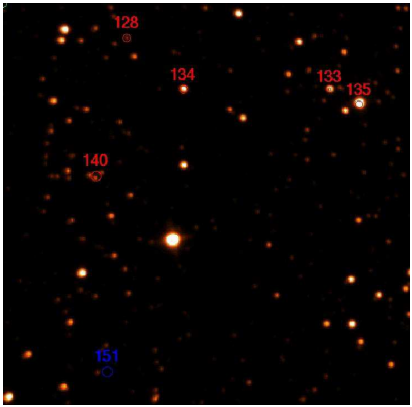,angle=0,width=5.4cm}
}
\caption{A SOFI $K_s$-band image of a part of field A2. Red circles
indicate positions of the detected soft X-ray sources, while the blue
one is a hard X-ray source.
}
\end{figure}

\end{document}